\documentclass[11pt,a4paper]{article}
\usepackage{jinstpub}

\usepackage{graphbox}
\usepackage[noabbrev]{cleveref}
\usepackage{natbib}

\usepackage{lineno}

\usepackage{subfiles} 

\usepackage{xspace}
\newcommand{\cpp}{\texttt{C++}\xspace}
\newcommand{\vxb}{\ensuremath{\vec{v} \times \vec{B}}\xspace}
\newcommand{\xmax}{\ensuremath{X_{\rm max}}\xspace}

\title{Framework and Tools for the Simulation and Analysis of the Radio Emission from Air Showers at IceCube}
\author[16]{R. Abbasi,}
\author[60]{M. Ackermann,}
\author[17]{J. Adams,}
\author[11]{J. A. Aguilar,}
\author[21]{M. Ahlers,}
\author[50]{M. Ahrens,}
\author[22]{J.M. Alameddine,}
\author[30]{A. A. Alves Jr.,}
\author[42]{N. M. Amin,}
\author[40]{K. Andeen,}
% ...
\affiliation[11]{Universit{\'e} Libre de Bruxelles, Science Faculty CP230, B-1050 Brussels, Belgium}
% ...
\affiliation[16]{Department of Physics, Loyola University Chicago, Chicago, IL 60660, USA}
\affiliation[17]{Dept. of Physics and Astronomy, University of Canterbury, Private Bag 4800, Christchurch, New Zealand}
% ...
\affiliation[21]{Niels Bohr Institute, University of Copenhagen, DK-2100 Copenhagen, Denmark}
\affiliation[22]{Dept. of Physics, TU Dortmund University, D-44221 Dortmund, Germany}
% ...
\affiliation[30]{Karlsruhe Institute of Technology, Institute for Astroparticle Physics, D-76021 Karlsruhe, Germany }
% ...
\affiliation[40]{Department of Physics, Marquette University, Milwaukee, WI, 53201, USA}
% ...
\affiliation[42]{Bartol Research Institute and Dept. of Physics and Astronomy, University of Delaware, Newark, DE 19716, USA}
% ...
\affiliation[50]{Oskar Klein Centre and Dept. of Physics, Stockholm University, SE-10691 Stockholm, Sweden}
% ...
\affiliation[60]{DESY, D-15738 Zeuthen, Germany}
% ...

\emailAdd{analysis@icecube.wisc.edu}
\date{\today}
\keywords{Air showers, radio frequency, simulation, cosmic rays}

\abstract{The Surface Enhancement of the IceTop air-shower array will include the addition of radio antennas and scintillator panels, co-located with the existing ice-Cherenkov tanks and covering an area of about 1\,km$^2$. Together, these will increase the sensitivity of the IceCube Neutrino Observatory to the electromagnetic and muonic components of cosmic-ray-induced air showers at the South Pole. The inclusion of the radio technique necessitates an expanded set of simulation and analysis tools to explore the radio-frequency emission from air showers in the 70\,MHz to 350\,MHz band. In this paper we describe the software modules that have been developed to work with time- and frequency-domain information within IceCube's existing software framework, IceTray, which is used by the entire IceCube collaboration. The software includes a method by which air-shower simulation, generated using CoREAS, can be reused via waveform interpolation, thus overcoming a significant computational hurdle in the field.
}

\notoc

\begin{document}


\maketitle

\section{Introduction}
\label{s:intro}

Cosmic rays with energies above 1\,PeV are studied using large (1\,km$^2$ to 1000\,km$^2$), sparse arrays of detectors on the ground that observe the \emph{air showers} that are produced in the interactions of the primary cosmic rays with nuclei in Earth's atmosphere, rather than directly observing the primary particle.
A major design consideration of current and future air-shower arrays includes the ability to directly measure the size of the electromagnetic and hadronic cascades for individual events.
A modern approach to do this involves combining various detector types to make complimentary measurements of the same air shower.

Radio antennas have proven to be particularly useful in this pursuit as they are a relatively cost-effective technology that can be used to make an accurate measurement of the development and size of the electromagnetic content in air showers~\cite{Buitink:2014eqa,PierreAuger:2015hbf,Bezyazeekov:2018yjw}.
Likewise, measurements of muons in air showers have been performed using particle detectors at cosmic ray observatories~\cite{PierreAuger:2014ucz,KASCADE-Grande:2017wfe,TelescopeArray:2018eph,PierreAuger:2020gxz,PierreAuger:2021qsd,IceCube:2021tuv}.
The combined knowledge of the electromagnetic and hadronic content in individual air showers can be used to provide a more accurate estimate of the mass and energy of the primary~\cite{PierreAuger:2016qzd,Holt:2019fnj}.

The IceTop Surface Enhancement~\cite{Schroder:2019suq} will feature 32 stations, each consisting of three SKALA-v2 antennas~\cite{SKALAV2_LNA} and eight 1.5\,m$^2$ scintillator panels~\cite{Oehler:2021qla}. The instrumentation will be installed within the footprint of IceTop, the existing 1\,km$^2$ array of ice Cherenkov detectors, located at the South Pole~\cite{IceCube:2012nn}.
Triggered by the scintillator panels, the antennas will be used to measure the atmospheric depth at which the electromagnetic content of an air shower reaches a maximum, \xmax, and to make a calorimetric measurement of the energy content of the electromagnetic cascades.
The radio emission from air showers is a result of two effects, the dominant geomagnetic effect and the subdominant Askaryan effect~\cite{Askaryan:1961pfb}.
The former arises when the positrons and electrons are deflected oppositely by the Lorentz force in the Earth's magnetic field while the latter is due to the gradual buildup of a net negative charge in the shower front during development.
This radiation will be sampled in the 70\,MHz--350\,MHz band and read out with 1\,ns resolution, amplified and digitized locally at each station, and then sent to a centralized location for analysis.
The information about the electromagnetic content will be combined with the observations from the scintillator panels and ice-Cherenkov tanks on the surface as well as with the measurements of $>$\,300\,GeV muons by the optical sensors that comprise the in-ice detectors of the IceCube Neutrino Observatory~\cite{IceCube:2016zyt}.
The addition of such information will make IceCube an essential experiment for studying the highest energy Galactic cosmic rays thanks to its capability to obtain a more complete picture of the particle content and development of individual air showers.

The inclusion of additional air-shower detection methods to IceCube requires the development of suitable software tools that can handle multi-detector simulations and reconstructions.
The current analysis framework used by the collaboration, IceTray~\cite{de2005icetray}, has been successful for the analysis of neutrino and cosmic ray observations using the existing surface and in-ice detectors.
In this paper we focus on the proposed radio extension, and to take full advantage of the capabilities that this addition brings, the IceTray framework has been extended for the analysis of the radio frequency (RF) information that will be recorded by the antennas on the surface of the ice at the South Pole.\footnote{For information about the simulation techniques for the scintillator panels, see~\cite{Leszczynska:2019ahq}.}
We briefly describe the existing analysis framework and detail the implementation of tools for the end-to-end simulation of the response of the antennas/hardware and the analysis of time- and frequency-domain information in \cref{s:sim_anal}.

As part of the standard simulation chain for radio events, we developed a method to re-use air shower simulations produced by the CoREAS~\cite{Huege:2013vt} extension of CORSIKA~\cite{corsika}.
Via this method, the radio emission, as given by CoREAS, is directly interpolated to produce the expected waveform at any other point in the radio footprint.
This technique can save orders of magnitude in total computing time when producing a library of simulated air showers and is described in \cref{s:waveform_interpolation}.


\section{Framework for Simulation and Analysis}
\label{s:sim_anal}

The general-purpose simulation and analysis framework developed by the IceCube Collaboration, IceTray, has been used to perform physics analyses on the data collected by the digital optical modules that comprise the photo-sensitive elements of the in-ice and surface components of the observatory~\cite{IceCube:2016zyt}.
The framework includes the flexibility to perform analyses that require the simultaneous measurements of air showers by the two distinct components of the observatory~\cite{IceCube:2019hmk,IceCube:2019scr,Bai:2019awp}.
The extension includes the data structures, analysis tools, relevant readers/writers, etc. for the analysis of an antenna array and naturally fits within this framework. 

In this section, we briefly describe the design and philosophy of IceTray and the extension for analysis of RF information in \cref{ss:IceTray}.
We then detail the modular tools that have been developed for simulation and analysis of the array of surface antennas in \cref{ss:modular_design}.
Finally, in \cref{ss:end_to_end_sim} we describe the method by which the end-to-end simulation is performed, including the response of all the readout components.

\subsection{Extension of the IceTray Design}
\label{ss:IceTray}

A key feature of the IceTray framework is the lack of a rigid, predefined data structure.
Instead, relevant pieces of data are stored in \emph{frames}.
Each frame contains a snapshot of the status of the observatory (detector locations and orientation, calibration information, etc.), the data read out from an observed/simulated event, and/or analysis information (e.g. reconstruction variables).
These generally correspond to a triggered readout of the relevant detectors.
The frame is implemented as a \cpp hash table, which allows for any number of pieces of data to be stored inside without specifying the data-type or how many pieces/types of data might be required by a user.
This allows for flexibility when designing an analysis, but, more relevant to this paper, also makes the framework easily extendable for handling RF information.

A critical extension for processing RF information is the representation of time-domain waveforms in the frequency domain.
For this, we implemented a data structure, which internally handles all discrete Fourier transformations, keeping track of the domain in which the data is most up to date and only requesting a transformation be performed when necessary.
This is particularly important in a module-based framework (see \cref{ss:modular_design}) where an analyzer may not know in which domain (i.e. time or Fourier) the information has been altered most recently by a particular algorithm.
Such an approach has been extensively used in the RF analysis code by other astroparticle experiments~\cite{PierreAuger:2011btp,Glaser:2019rxw}.
The underlying fast Fourier transformation (FFT) algorithm is handled by the open-source library, Fastest Fourier Transform in the West (FFTW3)~\cite{Frigo:2005zln}.
This library includes an optimization of the specific transformation algorithm based on the length of the waveform.
Two representations have been implemented for this data structure for use with 1D and 3D waveforms, typically used to represent the time series of voltages and electric field vectors, respectively.

\begin{figure}[t]
\centering
\includegraphics[align=c,width=0.48\textwidth]{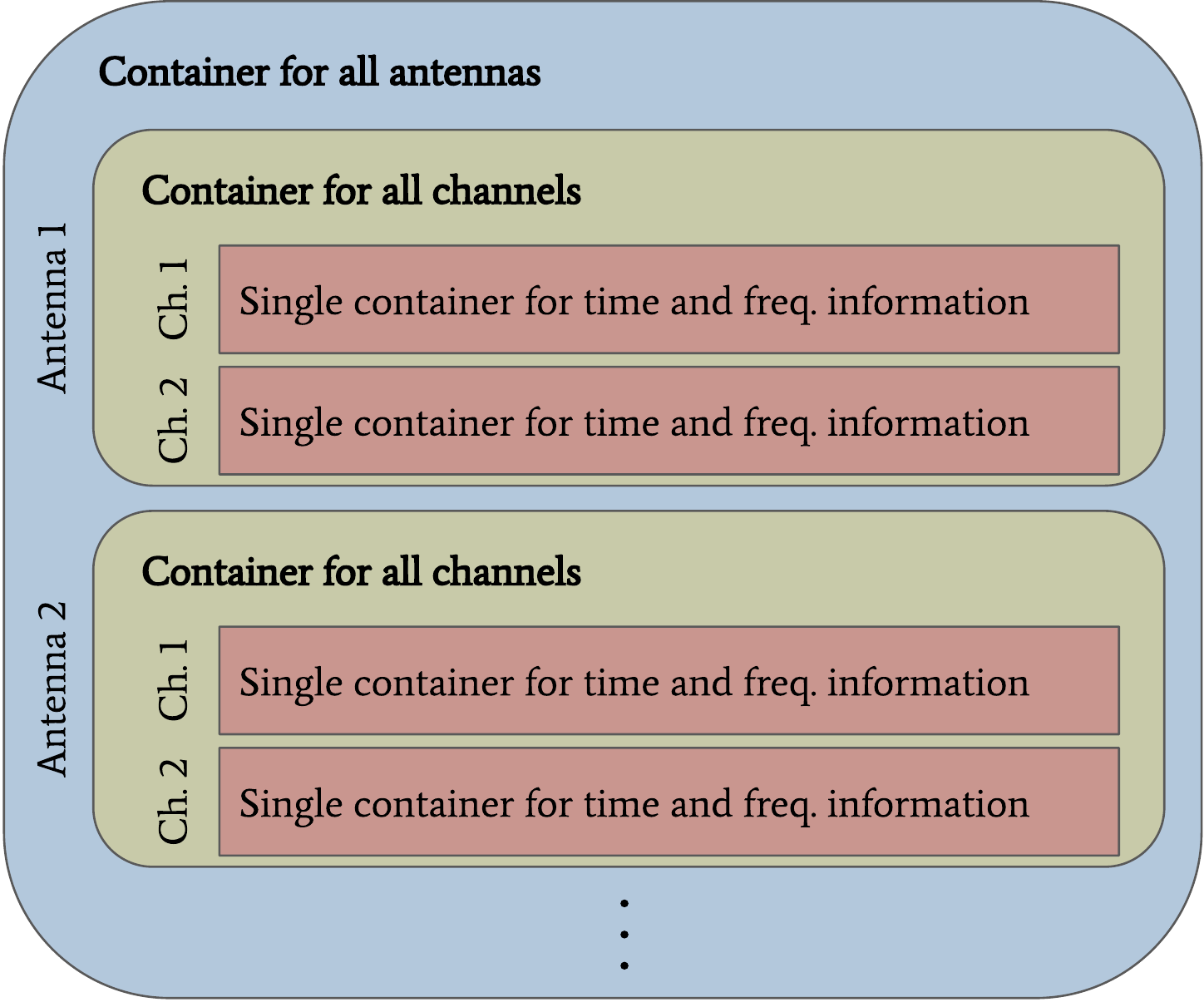}
\caption{The radio information is stored as a series of nested \cpp maps (represented by the blue/gold boxes). The outer map (blue) contains one entry for each antenna, while the inner maps (gold) contain an entry for each waveform associated with that antenna. The waveform information are stored in a format which facilitates the conversion to/from the Fourier domain (see text).
}
\label{fig:radio_data_diagram}
\end{figure}

To store RF waveform information, a specific class has been added that includes association between the antenna, its orthogonal arms (corresponding to two different \emph{channels}), and the corresponding waveforms.
Two nested \cpp maps hold the information for the radio array, shown schematically on the left side of \cref{fig:radio_data_diagram}.
Each antenna has an entry containing a map with all of the read-out information from the corresponding channels.
This map structure was chosen for flexibility so as to allow for future additions/modifications to the array, and the possibility to include antennas with more than two channels.
As with the rest of the IceTray framework, the same data structures are used for simulated and observed data so that each analysis can be developed on the former and directly applied to the latter.

The IceTray framework already includes tools for reading and writing the information contained within the frames~\cite{de2005icetray}.
A custom binary file structure has been developed by the collaboration for (de)serialization of the frames.
The code also includes methods by which data can be written in common, open-source formats, such as \texttt{hdf5}~\cite{folk2011overview} and \texttt{ROOT}~\cite{Brun:1997pa}.
The radio data structures, described above, have been integrated into these operations.
Additionally, a reader for CoREAS simulation output has been developed to parse the electric field information and relevant air shower properties.

\subsection{Modular Design within IceTray}
\label{ss:modular_design}

IceTray was designed with the intent for modular processing of observed and simulated data.
Within this framework, individual \emph{modules} are used to perform a specific, limited calculation/manipulation of the data.
To perform an analysis, the analyzer simply adds the relevant modules to a sequential list, called a \emph{tray}.
Each module then runs, in the order that they were added to the tray, on each frame.
Some modules are shared across the collaboration, for instance, the ones that propagate simulated leptons through the ice or those that accept/reject specific detectors in a particular event.
New shared modules have been developed to allow for the simulation and analysis of RF information for the radio antennas of the surface array.

As an extension to the tools to parse air-shower simulation output from CORSIKA, a module that also reads the corresponding CoREAS output has been included.
This reader works as a stand-alone procedure or in tandem with the previously-existing CORSIKA parser that is used for injecting secondary particles when simulating the response of the optical sensors and scintillator panels.
This is a crucial step to simulate the response of the entire observatory for a single generated air shower; see \cref{ss:end_to_end_sim}.

Several tools to perform commonly used calculations for time- and frequency-domain information are also included. These tools enable an analyzer to
\begin{enumerate}
    \item facilitate the up- and down-sampling of waveforms to a user-specified sampling rate;
    \item apply a band-pass filter for a desired transfer functions such as a box (top-hat in Fourier space) and Butterworth~\cite{butterworth1930theory};
    \item apply a phase gradient in Fourier space to produce a cyclic shift in the time-domain waveform.
\end{enumerate}

The inclusion of noise is particularly important in an RF analysis as it often sets the limit on the energy threshold and/or precision of an analysis.
For this purpose, two modules have been written.
The first generates uncorrelated noise and adds it to waveforms.
The dominant source of RF noise at the South Pole is due to the emission from Galactic and extragalactic sources and is incorporated into the code using the Cane model of the brightness, $B_{\rm Cane}(f)$~\cite{cane1979spectra}.
Uncorrelated noise is generated by first calculating the effective temperature,
\begin{equation}
    T_{\rm Cane}(f) = \frac{\lambda^2}{2 k_b} B_{\rm Cane}(f).
\end{equation}
The rms voltage for each frequency mode, $f_i$, is then given by,
\begin{equation}
    V_{\rm rms}(f_i) = \sqrt{\frac{Z\,k_b\,T_{\rm Cane}(f_i)\, A_{\rm eff}\, \Delta f}{2}},
    \label{eq:simulated_noise}
\end{equation}
where $k_b$ is the Boltzmann constant, $Z$ is the impedance of the system, $A_{\rm eff}$ is the integrated effective area, and $\Delta f$ is the width of one frequency bin.
Additional Johnson-Nyquist noise~\cite{Nyquist:1928zz}, with noise-temperature, $T$, can be added at each step in the processing chain, if desired,
\begin{equation}
    V_{\rm rms}(f_i) = \sqrt{Z\,k_b\,T\,\Delta f},
\end{equation}
For both types of noise, the voltage amplitudes, $V_{\rm rms}(f_i)$, are given a random phase and then transformed back into the time domain to produce incoherent, uncorrelated noise in the time domain.
The second module for generating noise directly adds waveforms of measured background in the time domain. This process uses snapshots of the local background using a fixed-rate trigger and can thus produce more realistic noise, which, in general, includes all the naturally occurring correlation between frequency modes and across antennas.

The final set of modules is used to (de)convolve the waveforms with the response functions of the antenna and the read-out electronics.
Two modules are also included that convert from/to electric fields to voltages, respectively, using the antenna response (see next section).
Likewise, the (un)folding of the response functions of the amplifiers, cables, and digitizers are handled by two dedicated modules.
All the response functions are loaded into memory as singleton objects, which can be queried by the modules during run-time, as several modules may require access to the same response function.
Further details on the method used to include these response functions will be given in the next section.


\subsection{Simulation of the End-to-End Response}
\label{ss:end_to_end_sim}

\begin{figure*}[t]
\centering
\includegraphics[width=0.9\textwidth]{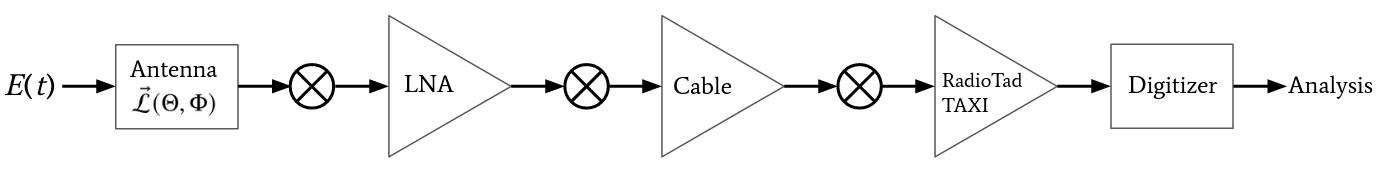}
\caption{The end-to-end simulation chain for impinging electric fields, $E$, includes the simulation of the several components. This includes the vector effective length of the antenna, $\mathcal{L}$, the low-noise amplifier (LNA), the cables, readout hardware, and the digitizer. The symbol $\bigotimes$ indicates where the complex-valued frequency-domain responses are included via a convolution.}
\label{fig:sim_schematic}
\end{figure*}

For a full simulation of the detector response, the gains and phases as a function of frequency must be taken into account for the deployed hardware.
In this section, we describe the method by which each of the detector components are included to perform an end-to-end simulation of a radio event, shown schematically in \cref{fig:sim_schematic}.

The vector effective length, $\vec{\mathcal{L}}$, of the SKALA-v2 antennas was simulated using CST Studio Suite~\cite{cstStudioStuite} to produce a table of the gain pattern and phase as a function of frequency and arrival direction in spherical coordinates, $\vec{\mathcal{L}}(f,\theta,\phi)$.
This was done in steps of 1\,MHz in frequency for 50\,MHz--350\,MHz and steps of 1$^\circ$ in arrival direction.
If needed, the values between those in the table are interpolated separately for the phase and the logarithm of the gain using bicubic-interpolation.

The voltage produced in an antenna from an impinging electric field, $\vec{E}$, is calculated in the Fourier domain using $V(f) = \vec{\mathcal{L}}(f, \theta, \phi) \cdot \vec{E}(f)$.
Note here that the relevant $\theta$ and $\phi$ are those of the Poynting vector, which are approximated using the propagation direction of the primary particle, accurate to $\simeq$\,1$^\circ$~\cite{Apel:2014usa}.

\begin{figure}[t]
\centering
\includegraphics[width=0.5\columnwidth]{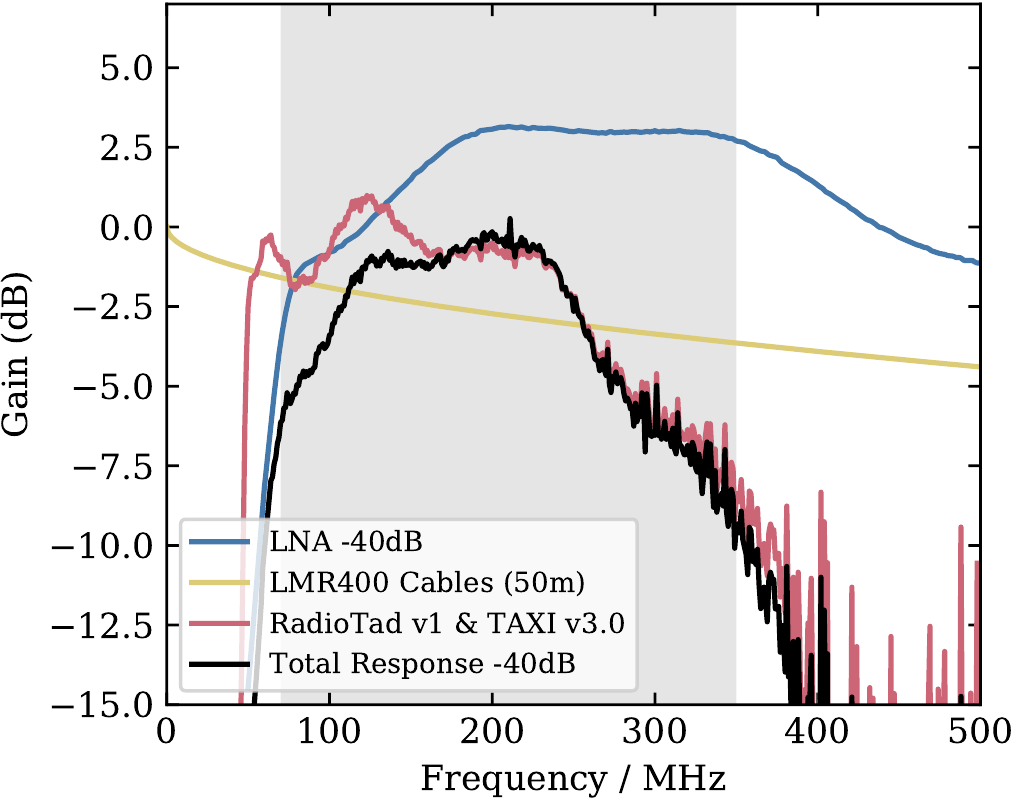}
\caption{The power gain of the LNA (blue) as well as the cables (gold) and readout electronics (red) are shown. The combined response is indicated in black. The gains of the LNA and the total have been shifted by $-40$\,dB for visual clarity. The grey band indicates the nominal frequency band of the system, 70\,MHz--350\,MHz.}
\label{fig:electronics}
\end{figure}

The remaining electronics are then included in a similar way, using interpolations of look-up tables for the corresponding gains and phases.
The response table for the low-noise amplifier (LNA) has been simulated using AWR Design Environment~\cite{arwDesign} and includes a roughly 40\,dB gain as shown in \cref{fig:electronics}.
The response of the 50\,m LMR400 coaxial cables has been measured directly in the lab as a function of temperature~\cite{Renschler:2020kcu}.
Finally, the combined response of the signal pre-processing board, RadioTad v1, and data acquisition module (TAXI v3.0) have also been measured in the lab~\cite{Oehler:2021qla}.
The combined gain pattern of the electronics, shown in \cref{fig:electronics}, results in a nominal frequency band for the system of 70\,MHz--350\,MHz.
The complex-valued responses, $R(f)$, are incorporated in the Fourier domain via a multiplication, $V'(f) = R(f)\times V(f)$, equivalent to a convolution of the time-domain waveforms, $V(t)$.

The final step of the end-to-end simulation, digitization, is the only operation performed in the time domain.
Here, the real-valued voltages are truncated to an integer corresponding to the 14-bit analog-to-digital converter (1\,V$_{\rm pp}$) on the TAXI board.

If desired, simulated noise can be added to the voltage waveforms at any step(s) in the amplification chain.
Otherwise, measured waveforms, which inherently include all the hardware responses, can be added directly after the digitization step.

\begin{figure}[t]
\centering
\includegraphics[width=1.0\columnwidth]{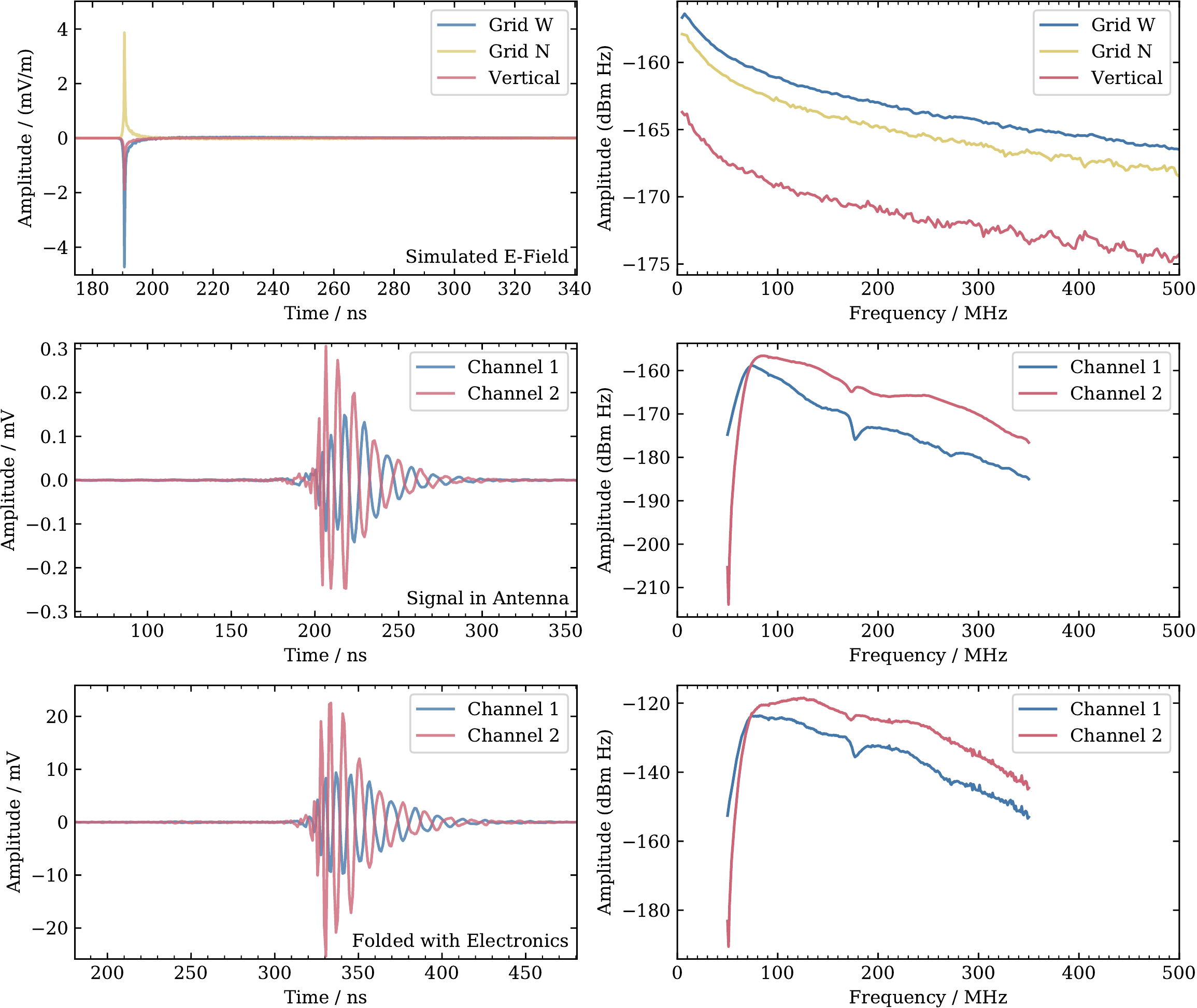}
\caption{The time series (left) and frequency spectra (right) are shown for three steps of the end-to-end processing of an electric field to a voltage. The top plots represent the raw electric field in the north, west, and vertical directions. The middle plots show the voltage produced in the channels of the crossed-dipole SKALA-v2 antenna. The bottom plot is again the same after convolving the response of the electronics described in \cref{ss:end_to_end_sim}.}
\label{fig:step_by_step}
\end{figure}

An example of the end-to-end simulation at three points during the processing are shown in \cref{fig:step_by_step}.
The left/right columns show the time series and corresponding frequency spectrum, respectively.
The top row includes the components of the electric field, transformed into the local geographic coordinate system.
The middle and bottom plots show the voltages after being convolved with the antenna response and the readout electronics (before digitization), respectively.
\section{Interpolation of CoREAS Simulations}
\label{s:waveform_interpolation}

The goal of the IceTop Surface Enhancement is to simultaneously measure air showers using various detection methods to disentangle the muonic and electromagnetic content and ultimately improve the estimation of the energy and mass of the primary particle.
To simulate the combined response of the multi-detector design, we generate CORSIKA/CoREAS simulations that simultaneously output the secondary particle content on the ground and the radio emission.
A major computational challenge in the analysis of any radio array is the order of magnitude increase in additional computation time to generate a single air shower simulation including radio emission.

The radio emission calculated in CoREAS uses the endpoint formalism in which the electric field that is generated by all the particles in an air shower is determined iteratively over many small steps~\cite{James:2010vm,Ludwig:2010pf}.
This must be calculated for each particle at each step steps for some number of locations in space, $\{\vec{r}_{{\rm ant},i}\}$, which must be specified \emph{before} beginning the CORSIKA/CoREAS simulation.
This has two main consequences.
Whereas a pure CORSIKA simulation, with the radio emission turned off, can take tens of minutes for a 30\,PeV air shower on current hardware, the inclusion of the radio calculation can take tens of hours and scales with the number of $\{\vec{r}_{{\rm ant},i}\}$.
Furthermore, the practice of \emph{resampling}, wherein the particles of a single air shower are injected into the detectors several times for different core locations, is generally not possible since the detector locations, with respect to the shower axis, have already been chosen.
We present a method that allows for the calculation of the expected waveform at an arbitrary point in the shower footprint by interpolating existing simulations where the electric fields are generated on a pre-defined pattern of locations.

Previous attempts have been made to use a star-shaped grid to estimate the scalar quantities describing the emission in the radio band~\cite{HoltDiplomaThesis2013}.
More recent attempts have included simulating the emission on such a grid and predicting the electric field waveforms at off-grid locations in space based on a semi-analytical model of the radio emission process~\cite{Butler:2017nvu}.
During the development of this work, a waveform interpolation technique, similar to the one described below, has been explored in~\cite{Tueros:2020buc}.
We note that the technique described there is not generally applicable for our use case as it assumes that \xmax can be treated as a point source.
However, at the South Pole, cosmic-ray primaries with energies above $\sim$\,$10^{17}$\,eV will regularly initiate air showers with an \xmax that is below the surface of the ice (altitude of 2840\,m a.s.l.)~\cite{TelescopeArray:2020bfv,Yushkov:2020nhr,Corstanje:2021kik}.

In this section, we describe a method by which the electric fields that are calculated by CoREAS can be interpolated.
In doing so, the location of the electric field at any point in the radio footprint can be determined, thus allowing for the resampling of the radio emission from air showers and also allowing template-based reconstruction techniques.
This method is thus a major component of the simulation chain of the antenna array and is also directly applicable to other experiments that detect air showers via the radio technique.

\subsection{Simulation on a Star-shaped Pattern}
\label{ss:SimOnStar}

\begin{figure}[t]
\centering
\includegraphics[align=c,width=0.8\columnwidth]{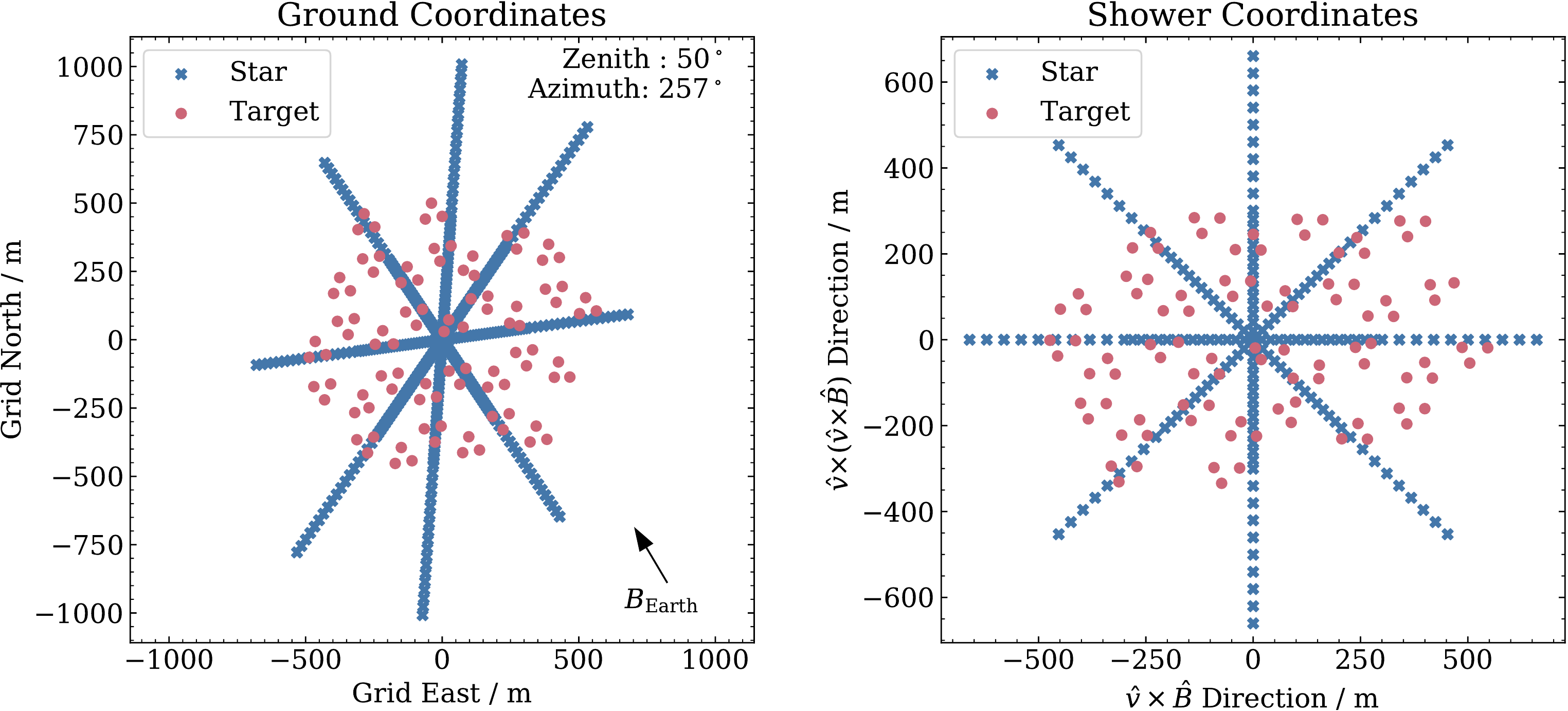}
\caption{Left: Locations of the simulated electric fields along the star-pattern (blue crosses) and test locations based on the planned array of antennas (red) in the local coordinate system at the South Pole. The projected direction of the local magnetic field is shown. Right: The same locations shown in the shower coordinate system with one axis along the direction of the Lorentz force.}
\label{fig:star_layout}
\end{figure}

The choice of pattern to use for the interpolation method is based on the symmetries of the radio emission in air showers.
The geomagnetic emission is linearly polarized along the Lorentz force, the \vxb direction, where $\hat{v}$ is the direction of shower propagation and $\vec{B}$ is the Earth's magnetic field.
The Askaryan emission is radially polarized with respect to the shower axis.
Depending on the polar angle about the shower axis, the interference of these two processes produces a net electric field that is not cylindrically symmetric.
Further, the coherent emission that is produced by the relativistic secondary particles generally can result in a Cherenkov ring on the ground~\cite{deVries:2011pa}.
As a result, the electric field strength does not change monotonically with distance from the shower axis.
To ensure that all of these features can be replicated properly, we chose a pattern of interpolation points in the shape of a star, with eight spokes that point radially outward from the shower axis, see \cref{fig:star_layout}.
Because of the polarization directions of the Askaryan and geomagnetic processes, there is a preferred orientation to analyze the total electric field emission.
For a given radius, the maximum emission will be where the interference is constructive and with a minimum where it is destructive, on the positive and negative \vxb sides of the shower axis, respectively.
It is important to ensure that both extrema are included in the sampling points that will be used for interpolation.
For this reason, two of the arms of the star are always parallel to the \vxb axis, as shown in the right panel of \cref{fig:star_layout}.

To develop and test the interpolation method, we created a library of simulated air showers using CORSIKA v7.7401 with the CoREAS radio extension.
We used proton and iron as primary particles and SIBYLL2.3d/FLUKA~\cite{Engel:2019dsg,Ferrari:2005zk} as the high/low energy hadronic interaction models.
The magnetic field corresponding to the South Pole, $54.6\,$\textmu T with a 17.8$^\circ$ zenith angle~\cite{chulliat2015us}, was used along with the average April South Pole atmosphere.\footnote{This atmosphere is implemented in CORSIKA as ATMOD 33.}  
Discrete zenith angles were chosen in steps of 5$^\circ$ from 0$^\circ$ to 65$^\circ$ whereas azimuth angles were chosen randomly.
For the verification of this method, we used simulations with a primary energy of 100\,PeV.

The electric field waveforms were generated using CoREAS with 0.2\,ns sampling for two sets of locations, simultaneously.
The first set of points were chosen according to the layout of the antennas that will be deployed as part of the IceTop Surface Enhancement~\cite{Schroder:2019suq}.
The second set of points were chosen along the eight spokes of the star, equally spaced around the shower axis.
The locations along each arm are separated by 20\,m for distances up to 300\,m from the center and 40\,m beyond that up to 750\,m.
One extra antenna is added at 0.5\,m from the shower axis to provide an inner interpolation node.
All chosen locations for both sets of points are at the same altitude of 2840\,m above sea level.

Note that since the set of points corresponding to the IceTop Surface Enhancement is fixed in ground coordinates, and that the star pattern is fixed in shower coordinates to the direction of \vxb, the relative location of both sets strongly depend on the shower direction.
Shower directions with increasingly large zenith angles will have a star pattern that is increasingly stretched along the ground plane.
The changing direction of \vxb with both the zenith and azimuth angle causes a relative rotation of the star pattern spokes with respect to the ground plane.
Thus, the choice of random azimuth angles ensures that, in the \vxb coordinate system, a sampling location corresponding to the IceTop Surface Enhancement can occur at any point within the star.

\subsection{Interpolation of Coherent Waveforms}
\label{ss:freq_domain_interp}

The interpolation of the electric field waveforms, $\vec{E}(t)$, begins with a rotation into the \vxb coordinate system, described above.
Each of the three components along the $\vec{v}$, \vxb, and $\vec{B}$ directions are interpolated separately.
In an air shower, the emission along the $\vec{v}$-direction is generally the smallest of these and thus any numerical noise or non-coherent emission from an air shower will have the least effect on this interpolation procedure.
The interpolation is performed separately on each of the three components, $E(t_i)$, and the respective Fourier amplitudes, $\widetilde{E}(f_j) \equiv \mathcal{A}_{j}$ for time-bin, $i$, and frequency-bin, $j$.

First, the start-time, $T_{0,k}$, of sampling location, $k$, is recorded.
This time is then shifted by the arrival time of a plane wave moving at the speed of light,
\begin{equation}
    \Delta T_{0,k} = T_{0,k} - (\Delta \vec{r}_k \cdot \hat{v}) / c.
    \label{eq:interp_start_time}
\end{equation}
Here, $\hat{v}$ is the velocity unit vector of the cosmic ray primary, and $\Delta \vec{r}_k$ is the corresponding location at which the electric field was calculated, with respect to the core.
The use of $\Delta T_{0,k}$ rather than $T_{0,k}$ ensures that all values of $\Delta T_{0,k}$ are of the same order of magnitude, which was ultimately found to produce more accurate results.

In the next two steps, the complex-valued, $\mathcal{A}_{jk}$ are decomposed into the real-valued logarithm of the magnitudes,
\begin{equation}
    \mathcal{M}_{jk} = \log |\mathcal{A}_{jk}|,
    \label{eq:interp_amplitude}
\end{equation}
and a complex-valued unit vector representing the phase of $\mathcal{A}_{jk}$,
\begin{equation}
    \mathcal{P}_{jk} = \exp^{i \phi_{jk}} = \mathcal{A}_{jk} / |\mathcal{A}_{jk}|.
    \label{eq:interp_phase}
\end{equation}
The use of the unit vector in the complex plane, rather than directly calculating the phase angle, $\phi_{jk}$, was done to more naturally avoid the issues of interpolating cyclically-valued quantities.

The corresponding waveform for an arbitrary point within the simulated radio footprint is calculated using the $\Delta T_{0,k}$, $\mathcal{M}_{jk}$, and $\mathcal{P}_{jk}$.
\begin{figure}[t]
\centering
\includegraphics[align=c,width=0.4\columnwidth]{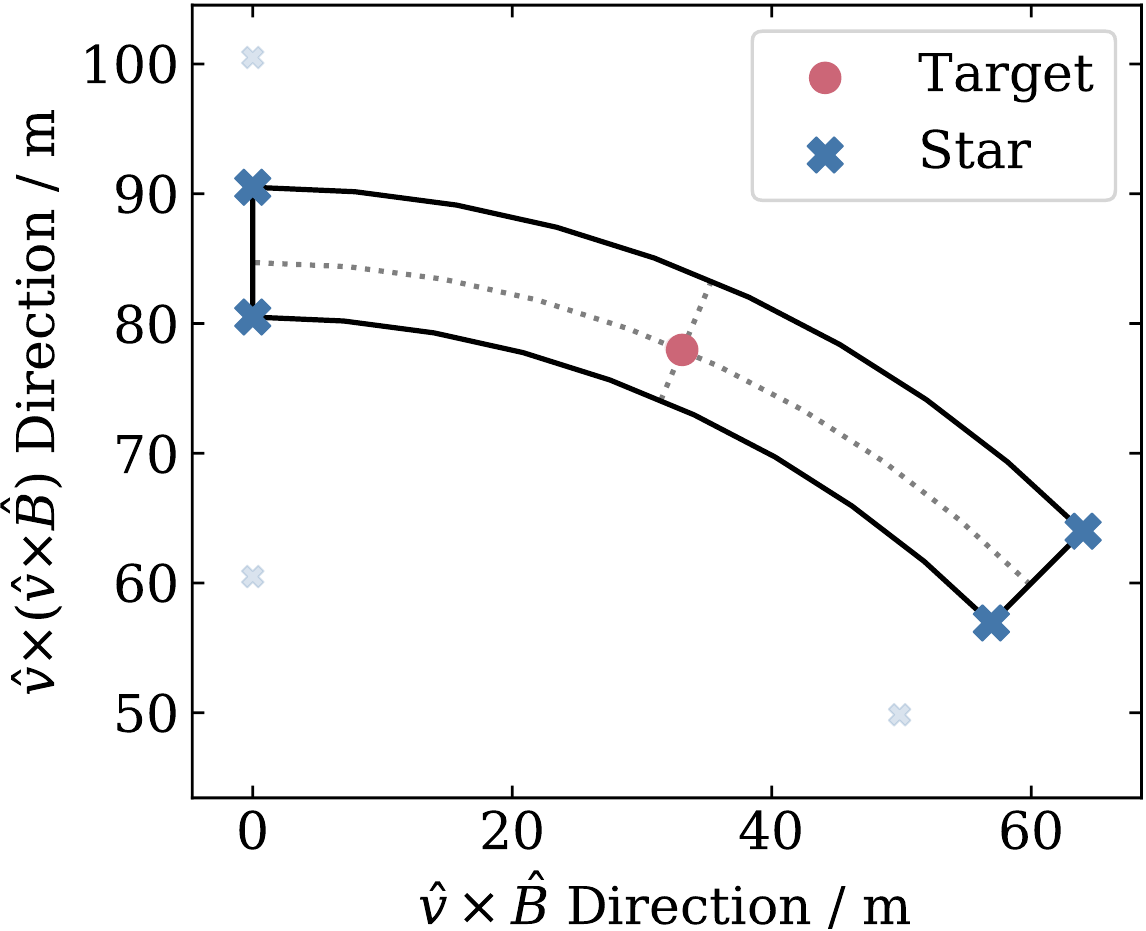}
\hspace{0.5cm}
\includegraphics[align=c,width=0.4\columnwidth]{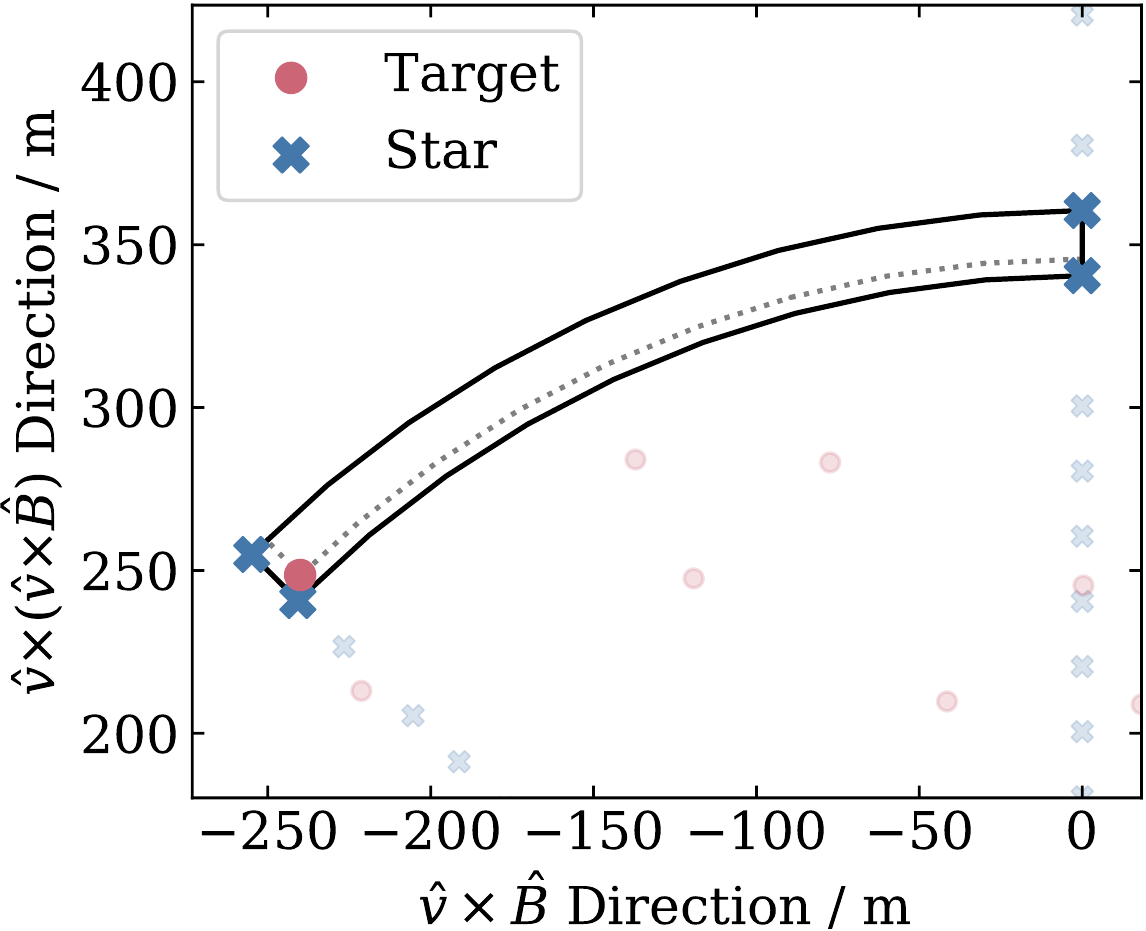}
\caption{Each panel is an example of the interpolation process to attain the expected waveform at a target location (red circle), using the four surrounding neighbors (blue crosses). The other target and star-pattern locations are displayed in a lighter color. 
These examples correspond to the simulated shower shown in \cref{fig:star_layout} and the interpolated waveforms in \cref{fig:interp_waveform}.}
\label{fig:nearest_neighbors}
\end{figure}
As shown in \cref{fig:nearest_neighbors}, the electric field at a given location is determined using the four nearest and surrounding locations on the simulated spokes of the star.
The corresponding values of $\Delta T_{0}$, $\mathcal{M}$, and $\mathcal{P}$ for the requested location are determined via a bilinear interpolation in polar coordinates, $r$ and $\theta$.
For $\Delta T_{0}$ and $\mathcal{M}$, this is a relatively straight forward process.
However, the interpolated value of $\mathcal{P}$ must be renormalized since, in general, the resulting value has a magnitude less than unity.
The frequency spectrum for the target location is constructed by inverting the decomposition process described in \cref{eq:interp_amplitude,eq:interp_phase}.
Finally, the time-domain electric field is recovered via a Fourier transformation and a start time is assigned to the waveform by inverting \cref{eq:interp_start_time}.

\subsection{Accuracy of the Interpolation Method}
\label{ss:interp_accuracy}
\begin{figure*}[t]
\centering
\includegraphics[width=0.45\textwidth]{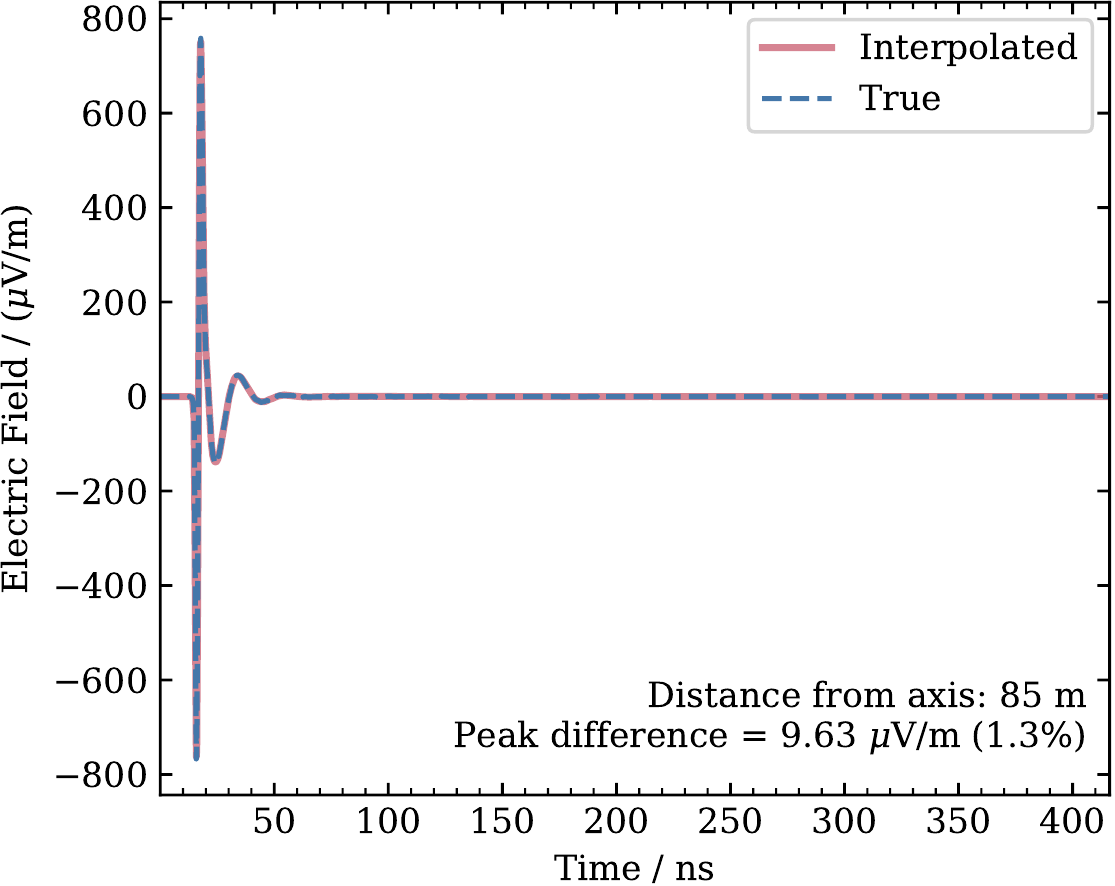}
\hfill
\includegraphics[width=0.43\textwidth]{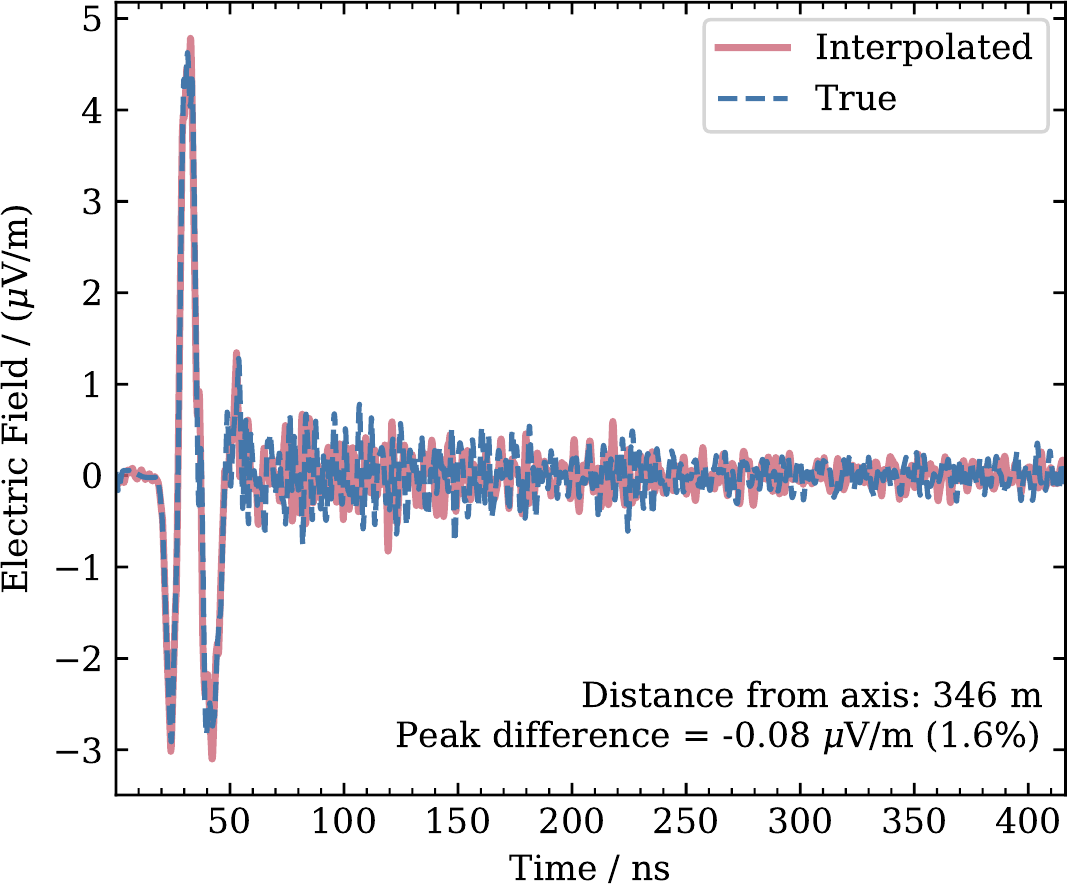}\\
\vspace{0.2cm}
\includegraphics[width=0.45\textwidth]{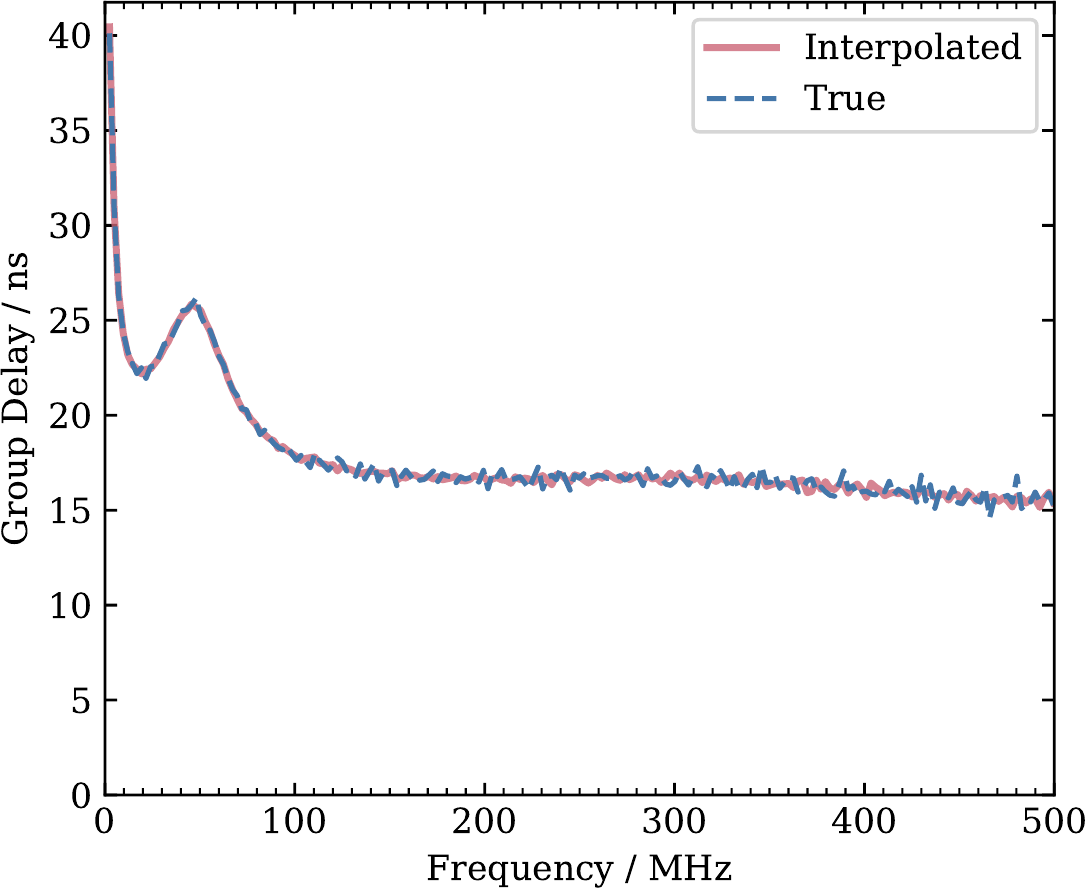}
\hfill
\includegraphics[width=0.45\textwidth]{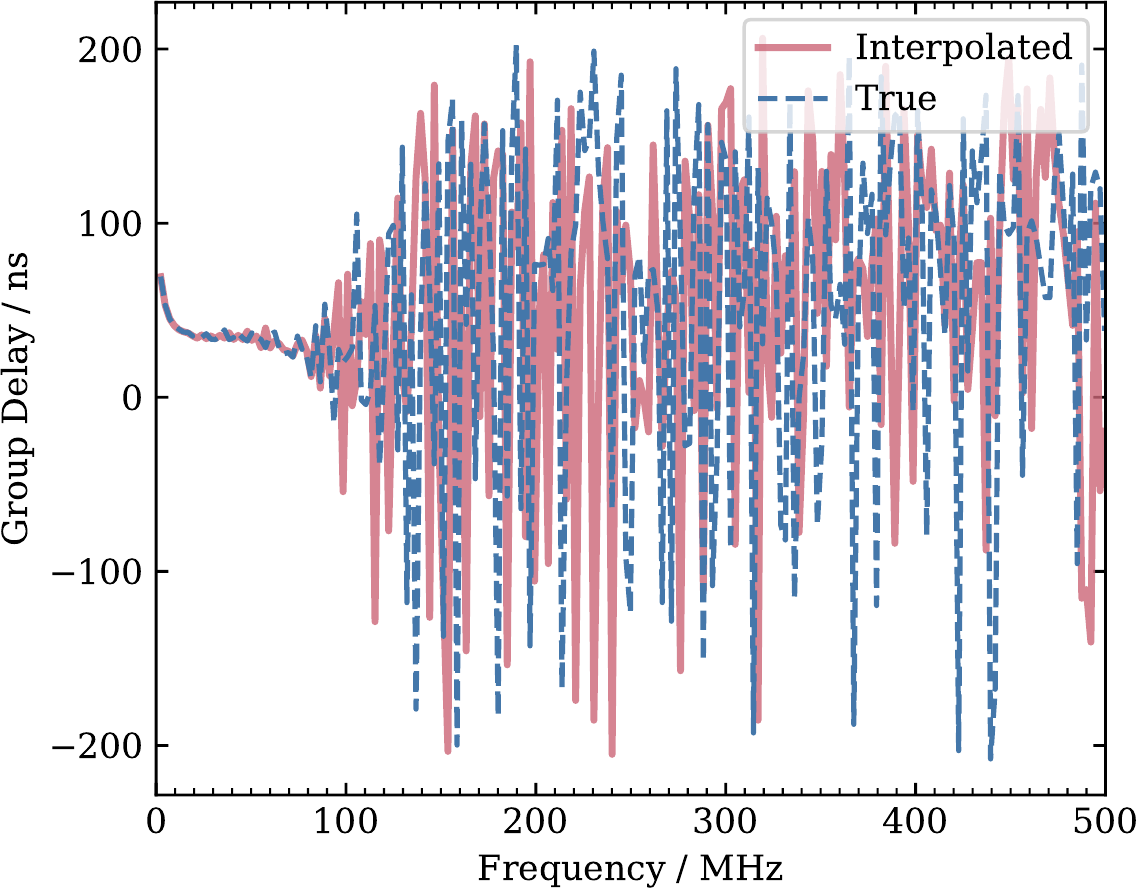}\\
\vspace{0.2cm}
\includegraphics[width=0.45\textwidth]{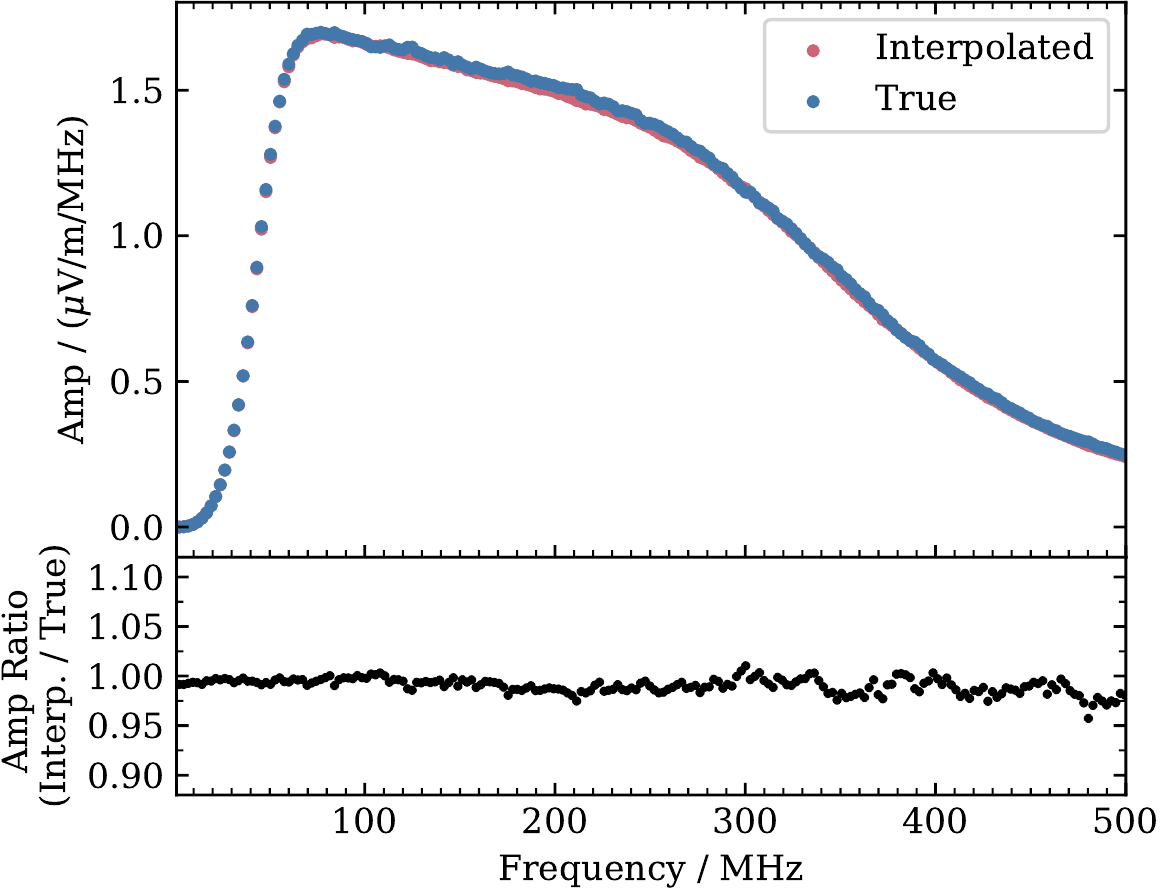}
\hfill
\includegraphics[width=0.45\textwidth]{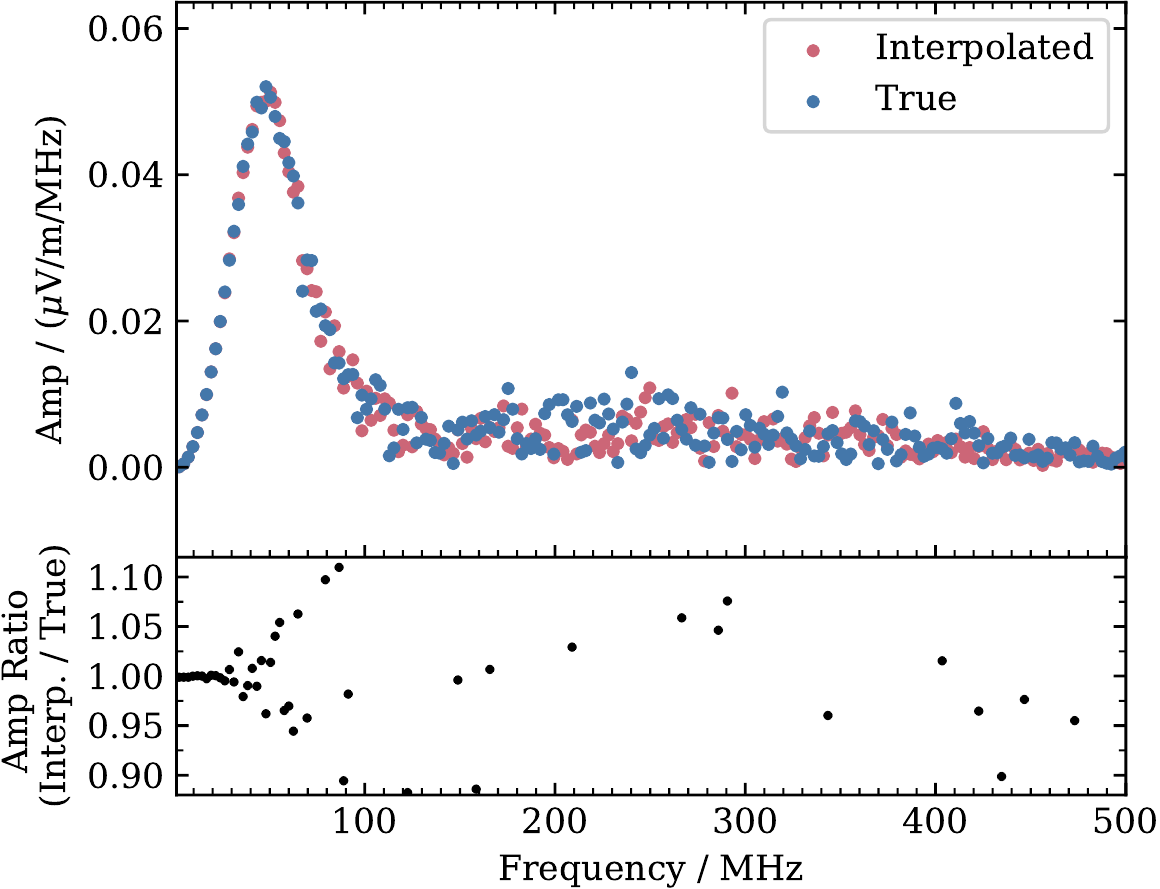}
\caption{The results of the interpolation method for two example locations that are closer/further from the shower axis (left/right columns). The plots in the top row show the time series for the exact CoREAS output (dashed blue) as compared to the interpolated values (solid red) at that location. The figures in the center and bottom rows show the group delay and Fourier amplitudes, respectively, for the true and interpolated values. The residuals of the Fourier amplitudes are given in the lower panels. In all plots, a third-order Butterworth filter has been applied for the band 50\,MHz--350\,MHz. These two examples correspond to the respective panels in \cref{fig:nearest_neighbors}.}
\label{fig:interp_waveform}
\end{figure*}

For understanding the power of the interpolation method, it is useful to study two examples of the interpolation procedure.
These are shown in \cref{fig:interp_waveform} for locations at 85\,m and 346\,m from the shower axis.
For the location closer to the shower axis, the coherence extends across the entire frequency band being studied, and the Fourier amplitudes are accurately described to within 3\%.
For the location further from the shower axis, there is coherence only below $\simeq$\,100\,MHz.

The deviation of the interpolated and true Fourier amplitudes above 100\,MHz in the second example is an expected behavior.
For radii on or inside the Cherenkov ring (located at $\simeq$\,100\,m for this shower), the underlying frequency spectra of the radio emission are relatively flat up to $\sim$\,0.1\,GHz--1\,GHz.
Likewise, the coherent emission from air showers exhibits a smooth and continuous evolution of the phase angle, as seen in the center-left plot of \cref{fig:interp_waveform}.
At radii outside of the Cherenkov cone, the frequency spectrum is steeper and includes a cutoff within the band of interest.
Beyond this cutoff, there is noise, which may be partially physical in origin, e.g. due to the non-coherent summation of emission from distant regions of the shower front.
However, artifacts can be introduced via the thinning algorithm in CORSIKA wherein the combination of particles into a single, weighted entity effectively creates perfectly coherent emission, locally.
As the interpolation method implicitly requires that the values being described are well behaved between the interpolation nodes, the phases and Fourier amplitudes cannot be replicated for non-coherent frequency modes.
However, for our use case, a high-accuracy reproduction of the non-coherent aspects of the signal is not important since such low-level noise would be dominated by other backgrounds in almost all cases.

\begin{figure}[t]
\centering
\includegraphics[width=0.45\columnwidth]{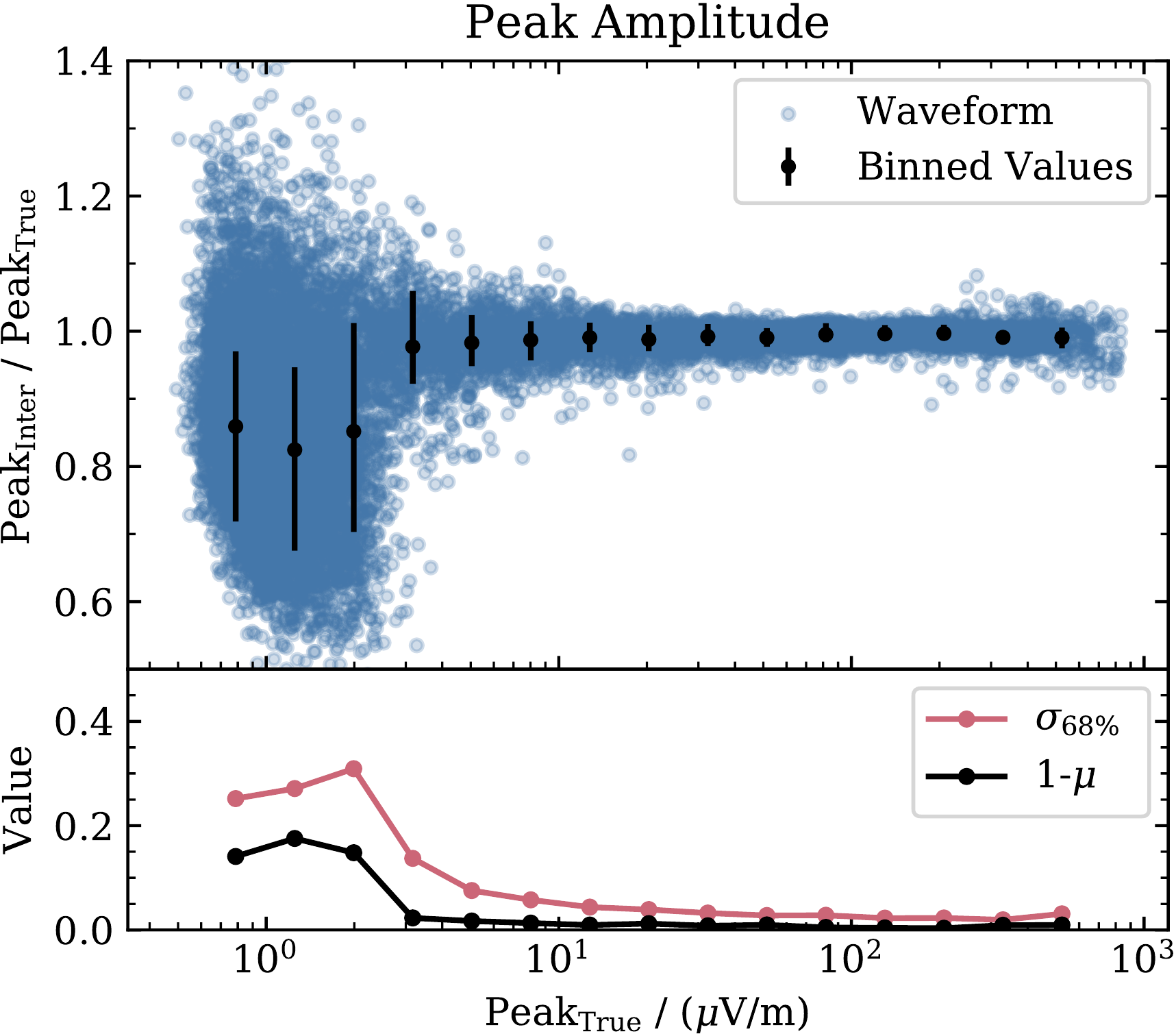}
\qquad
\includegraphics[width=0.45\columnwidth]{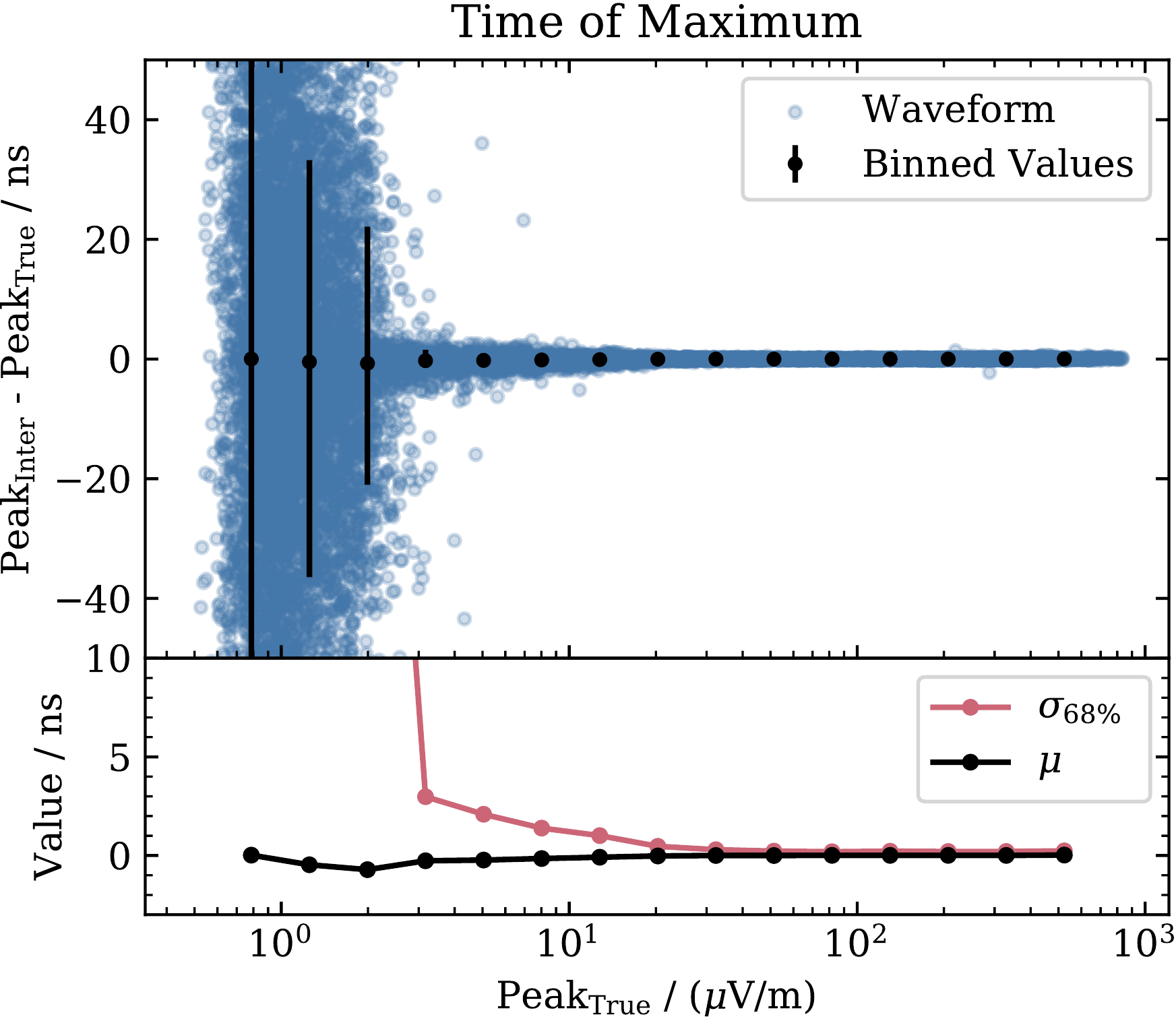}
\caption{Metrics for the accuracy and precision of the interpolation method are shown above for the Hilbert peak amplitude (left) and time-at-maximum (right). The values for individual waveforms are shown in blue. The median value is overlaid in black with error bars representing the 68\% interval. The bottom panels show the full width of the 68\% interval, $\sigma_{68\%}$, in red and the median value, $\mu$, in black.}
\label{fig:interp_results}
\end{figure}

We performed a statistical comparison of two quantities, chosen for their use in air-shower reconstructions~\cite{Tunka-Rex:2015zsa,Kostunin:2015taa}, to determine the overall accuracy of the interpolation method.
For each of the simulated locations on the proposed array, the amplitude- and time-at-maximum of the Hilbert envelope were calculated.
A comparison of these values for the true- and interpolated-pulses are shown in \cref{fig:interp_results}.
This study highlights two physical domains in which the interpolation method produces (in)accurate results.
The first domain, in which the frequency band of interest is either fully or mostly non-coherent, is characterized by signals with a peak amplitude below a threshold of $V_{\rm th}\simeq 3$\,$\mu$V/m.
Such signals occur far from the shower axis where spectral cutoff occurs at frequencies $\lesssim 50$\,MHz.
In this domain, the resolution is 10\% to 20\% in peak amplitude and a few tens of ns in the peak location.\footnote{The 68\% confidence interval should be compared to $2\sigma$ of a symmetric distribution.}
In the second domain, the waveforms are fully or mostly coherent, and the method works well (see example in the left panels of \cref{fig:interp_waveform}).
The resolution for waveforms with a peak amplitude near $V_{\rm th}$ is about 7\% in magnitude and 2\,ns in peak time.
For the largest signals shown here, the resolution is better than 1.5\% in amplitude with a difference in peak time that is comparable to the time binning of the simulated electric field (0.2\,ns). 
The example in the right panels of \cref{fig:interp_waveform}, which includes coherence only in a fraction of the band of interest, is indeed above $V_{\rm th}$, and has a peak amplitude of 4.6\,$\mu$V/m.

It is important to consider the way that this method scales with cosmic-ray energy.
The accuracy of the interpolation and the value of $V_{\rm th}$ is not tied to an absolute scale in field strength.
All simulations studied in this work include an amount of non-coherent emission that is proportional to the energy of the primary particle.
For cosmic rays with higher/lower energies, the distributions shift to larger/smaller values with the overall structure being unaffected and $V_{\rm th}/E_{\rm primary}$ being constant.
This has been tested using two additional sets of simulations with primary energies of 50\,PeV and 500\,PeV.

It is unknown what fraction of this non-coherent emission would actually occur in air showers as opposed to, for instance, numerical rounding effects or the extra coherence introduced by the thinning algorithm implemented in CORSIKA.
Thus, the systematic uncertainties introduced via the interpolation method should be taken into account when considering high energy simulations where the experimental background noise is comparable to $V_{\rm th}$.
Generally, this would need to be studied for individual experiments and frequency bands.

The metrics that were chosen to quantify the accuracy of this method are based on macroscopic observables of individual waveforms.
However, they do not necessarily encapsulate more low-level details, such as the exact shape of the waveform.
For analyses that are sensitive to the exact structure of the waveforms, such as using interferometry to reconstruct air showers~\cite{Romero-Wolf:2014pua,Schoorlemmer:2020low, LOPES:2021ipp,Schluter:2021egm}, a more detailed understanding of this method's limitations may be required.
In particular, the interferometry technique has been shown to require $<$\,1\,ns timing accuracy for the frequency band that will be used for the IceTop Surface Enhancement~\cite{Schluter:2021egm}.

We studied the impact of using different configurations of the star layout.
Using fewer than eight spokes produced large discrepancies in the peak amplitude of several tens of percent.
We also studied using an increased density of nodes along each spoke, up to one node every 5\,m.
For increased densities of nodes, the quality metrics shown in \cref{fig:interp_results} did not improve.
However, using a spacing between nodes larger than 20\,m produced less accurate results.


\subsection{Interpolation Example for an Observed Air Shower}

\begin{figure*}[t]
\centering
\includegraphics[width=0.9\columnwidth]{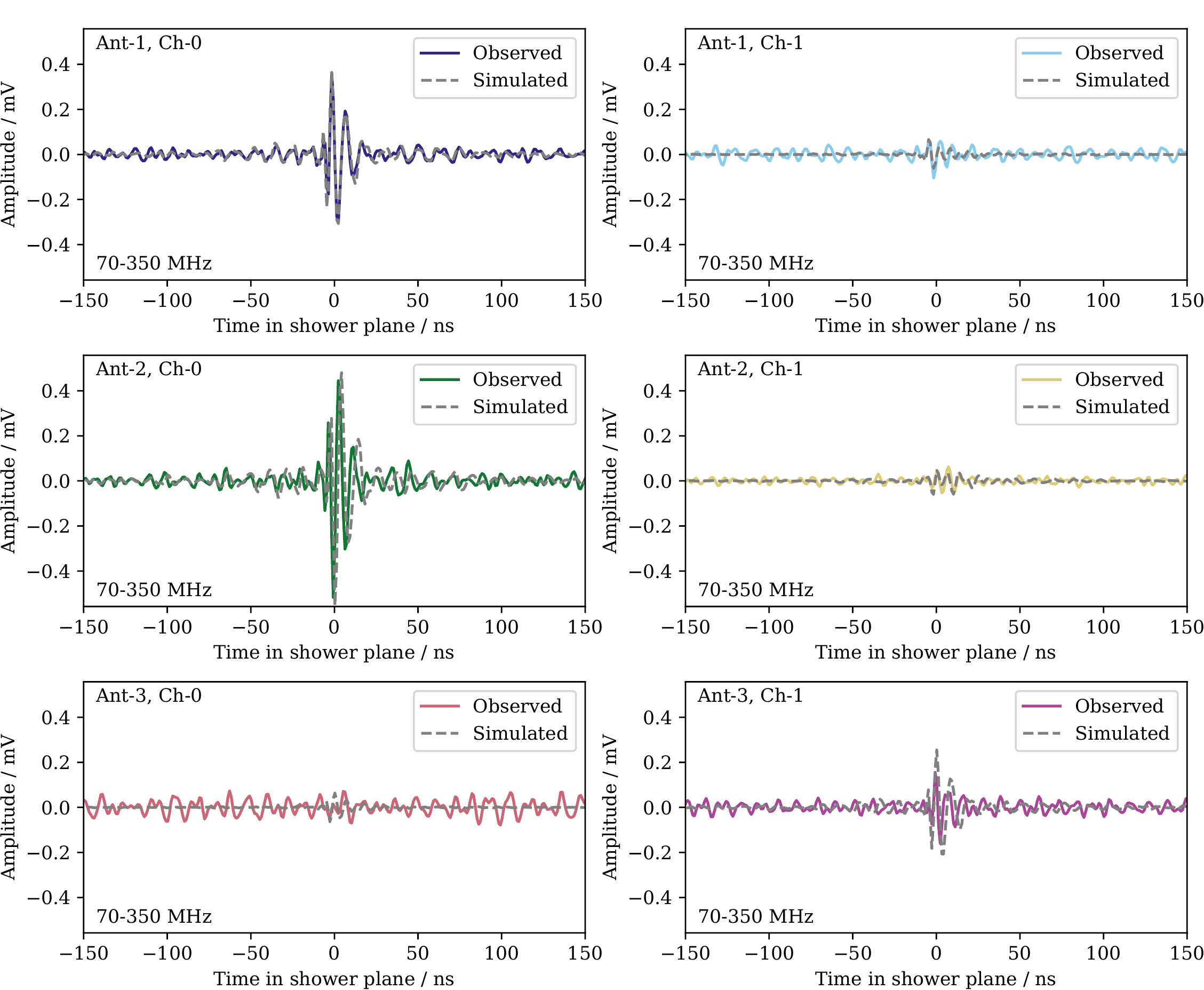}
\caption{The observed voltages, after deconvolving the response of the LNA, cables, and readout hardware, in the six arms of the prototype station at the South Pole are shown in solid color. The respective interpolated waveforms using a CoREAS simulation with a direction and core motivated by the IceTop reconstruction ($\theta = 32^\circ$, $\phi = 275^\circ, E_{\rm IT}=240$\,PeV) are overlaid in dashed gray. 
}
\label{fig:real_event}
\end{figure*}

An example of the combination of the tools described in this paper is shown in \cref{fig:real_event}.
We show the waveforms corresponding to the two channels of the three antennas for an air shower that was observed by the prototype station at the South Pole.
We simulated a proton shower using the star pattern with initial conditions based on the direction and energy estimated using the IceTop reconstruction~\cite{IceCube:2012nn}.
Using the core location from the IceTop reconstruction algorithm, the waveforms were interpolated for each antenna location and then folded with the antenna response.
The resulting voltages are overlaid in dashed gray.
General agreement is seen between the measured and observed voltages for this shower.
\section{Conclusions}
\label{s:conclusion}

In this paper, we detailed the tools developed by the IceCube Collaboration for simulating and analyzing the radio emission from air showers.
Within the observatory, the radio technique is unique given that it necessitates analyzing waveforms in the frequency domain as well as the time domain.
We described in \cref{s:sim_anal} the data structures that were included into the IceTray framework to enable this.
Building on the modular structure, several tools have been included to perform common RF calculations, such as band pass filtering and waveform resampling.

We described the way that the various aspects of the readout hardware are included into the analysis chain.
These take into account the gains and group delays that are associated with the antenna and the corresponding hardware.
Together, we are able to produce end-to-end simulations of the way that our detector responds to impinging electric fields, specifically those from CORSIKA/CoREAS simulations.
This also includes the addition of either non-coherent modeled noise or measured background waveforms from the antennas at the South Pole.

As an integral part of the simulation chain, we have developed a method whereby the electric fields produced in simulated air-showers can be interpolated.
This method is based on the interpolation of the Fourier components and the start time of the waveform to ultimately produce a time-domain electric field.
The interpolation method described above works well for the coherent emission that radio arrays typically measure.
For all studied energies, the peak amplitude and peak time of the air shower pulses can be faithfully reproduced to 5\% and 1.5\,ns, respectively, for waveforms with coherent emission in the band of interest.

While this tool was developed as part of our simulation scheme, it has wider applications for other experiments as well.
Since the radio technique is only viable for primary energies above $\sim$\,10\,PeV, the computational burden of producing a large library of air shower simulations is substantial.
Being able to resample air shower core locations when performing a detector simulation directly reduces the amount of time and resources needed to enable an analysis.
Additionally, as existing astroparticle experiments consider various extensions of antenna arrays, such as IceCube-Gen2~\cite{IceCube-Gen2:2020qha}, this method enables the production of a single air-shower library that can be used for testing multiple arrangements of antennas.
The second application for this method is in regards to the reconstruction of air showers.
One of the most precise methods for reconstructing the electromagnetic energy and \xmax is to produce many simulations for each observed event and to compare the observed and simulated waveforms directly~\cite{Bezyazeekov:2018yjw,Corstanje:2021kik,Mulrey:2020oqe}.
These simulations have initial conditions such as direction and core location based on the reconstruction of another detector (typically scintillator panels or Cherenkov tanks).
By interpolating the expected signal, one can also leave the core location as a free (or constrained) parameter while template matching to improve the accuracy of the reconstruction.


The framework and simulation tools that are described above have been included in the IceCube software repository.
With these, we can simulate the response of the entire observatory to air showers.
The combined detection technique is a necessary step to improve the precision of the estimation of the primary particle, specifically in the assignment of mass and energy.
We expect that the radio technique will play an important role in this regard with respect to the IceTop Surface Enhancement as well as for the proposed surface array of IceCube-Gen2~\cite{IceCube-Gen2:2020qha}.


\acknowledgments
The IceCube collaboration acknowledges the significant contributions to this manuscript from Alan Coleman, Abdul Rehman and Frank Schroeder.

We would like to thank Eloy de Lera Acedo for providing the simulated responses for the SKALA-v2 antenna and the LNA.

USA {\textendash} U.S. National Science Foundation-Office of Polar Programs,
U.S. National Science Foundation-Physics Division,
U.S. National Science Foundation-EPSCoR,
Wisconsin Alumni Research Foundation,
Center for High Throughput Computing (CHTC) at the University of Wisconsin{\textendash}Madison,
Open Science Grid (OSG),
Extreme Science and Engineering Discovery Environment (XSEDE),
Frontera computing project at the Texas Advanced Computing Center,
U.S. Department of Energy-National Energy Research Scientific Computing Center,
Particle astrophysics research computing center at the University of Maryland,
Institute for Cyber-Enabled Research at Michigan State University,
and Astroparticle physics computational facility at Marquette University;
Belgium {\textendash} Funds for Scientific Research (FRS-FNRS and FWO),
FWO Odysseus and Big Science programmes,
and Belgian Federal Science Policy Office (Belspo);
Germany {\textendash} Bundesministerium f{\"u}r Bildung und Forschung (BMBF),
Deutsche Forschungsgemeinschaft (DFG),
Helmholtz Alliance for Astroparticle Physics (HAP),
Initiative and Networking Fund of the Helmholtz Association,
Deutsches Elektronen Synchrotron (DESY),
and High Performance Computing cluster of the RWTH Aachen;
Sweden {\textendash} Swedish Research Council,
Swedish Polar Research Secretariat,
Swedish National Infrastructure for Computing (SNIC),
and Knut and Alice Wallenberg Foundation;
Australia {\textendash} Australian Research Council;
Canada {\textendash} Natural Sciences and Engineering Research Council of Canada,
Calcul Qu{\'e}bec, Compute Ontario, Canada Foundation for Innovation, WestGrid, and Compute Canada;
Denmark {\textendash} Villum Fonden and Carlsberg Foundation;
New Zealand {\textendash} Marsden Fund;
Japan {\textendash} Japan Society for Promotion of Science (JSPS)
and Institute for Global Prominent Research (IGPR) of Chiba University;
Korea {\textendash} National Research Foundation of Korea (NRF);
Switzerland {\textendash} Swiss National Science Foundation (SNSF);
United Kingdom {\textendash} Department of Physics, University of Oxford;
European Union – European Research Council, Horizon 2020.

\bibliographystyle{JHEP}
\bibliography{main}

\end{document}